\documentclass[a4paper,11pt]{article}

\usepackage[utf8]{inputenc}
\usepackage{newunicodechar}
\newunicodechar{→}{$\rightarrow$}
\usepackage{comment}
\usepackage{amsmath}
\usepackage{amssymb}
\usepackage{amsthm}
\usepackage{array}
\usepackage{booktabs}
\usepackage[singlelinecheck=off,justification=centering]{caption}
\usepackage{geometry}
\usepackage{graphicx}
\usepackage{float}
\usepackage{hyperref}
\usepackage{hyphenat}
\usepackage{longtable}
\usepackage{lscape}

\usepackage{natbib}

\usepackage{geometry}      
\usepackage{setspace}
\usepackage{tabularx}
\hypersetup{colorlinks,citecolor=blue,urlcolor=blue,linkcolor=blue}
\usepackage{soul,color}
\usepackage{threeparttablex}
\usepackage{url}
\usepackage{eurosym}
\usepackage{subcaption}

\newcolumntype{C}[1]{>{\centering\arraybackslash\hspace{0pt}}p{#1}}

\usepackage{tabularx,booktabs}
\newcolumntype{Y}{>{\centering\arraybackslash}X}

\usepackage{tabularx} 
\usepackage{array} 
\usepackage{rotating}

\newgeometry{top=2.7cm,bottom=2.7cm,left=2.7cm,right=2.7cm}

\newcommand{\sym}[1]{#1}

\title{What is the public’s social welfare function?\thanks{\footnotesize We thank Tim Besley, Francesco Capozza, Daniel Chandler, Crispin Cooper, Stephen Jenkins, Caspar Kaiser, Christian Krekel and Sara MacLennan for their comments and suggestions. We are grateful for financial support from the LSE Programme for Cohesive Capitalism. We thank YouGov for conducting the fieldwork. YouGov is not responsible for the analysis or interpretation of the data; any errors or conclusions are the authors' own.}}

\author{Richard Layard \\ LSE  \and Ekaterina Oparina\footnote{Corresponding author, email:  e.oparina@lse.ac.uk} \\ LSE}

\date{May, 2026}

\begin{document}

\maketitle
\thispagestyle{empty}

\begin{abstract}

\singlespacing{\noindent Optimal public policy requires a social welfare function defined over individual utilities. While there is substantial research on income-based social welfare functions, no published study has directly elicited public preferences over utility when measured by subjective wellbeing. Using a novel survey instrument with a representative UK sample (N=2,068), we estimate the public’s social welfare function for life satisfaction. We find significant aversion to wellbeing inequality, with a median isoelastic parameter $\alpha$=0.48. This implies a social welfare function approximately equal to the sum of square roots of individual utilities. The median respondent values improving the wellbeing of the least satisfied by one unit roughly twice as much as improving the most satisfied by one unit. Our findings provide ethically grounded distributional weights for wellbeing policy evaluation and cost-benefit analysis.} \\ \\

\noindent{\textbf{Keywords:}  social welfare function, inequality aversion, wellbeing, life satisfaction, distributional weights}

\end{abstract}

\doublespacing
\newpage

\section{Introduction}
\label{introduction}

We can only analyse optimal public policy if we have a social welfare function, defined over the vector of individual utilities (see, for example, \cite{atkinson_lectures_1980}).  Yet, surprisingly, there has been almost no research on what the public believes the form of such social welfare function should be.\footnote{The only exception is \citet{cooper_individual_2026}, who elicit wellbeing inequality aversion using gambles. We compare our approach with theirs below.}

There has been research over a different social welfare function, defined over the vector of individual incomes. But incomes and `willingness to pay' cannot reflect the impact on utility of being sick or sacked.\footnote{Willingness to pay can be computed where there is evidence from revealed preference. But this requires observed choices. The other approach, stated preference, is subject to known problems such as scope insensitivity and the dependence of the estimates on the type of question asked \cite{kahneman_economic_2000}.} The solution is to measure utility itself. The most commonly used measure is life satisfaction, where the question is `Overall, how satisfied are you with your life these days?' (0 to 10 where \textit{0 = Not at all}, \textit{10 = Completely satisfied}). Almost 90\% of OECD governments now ask this question (\cite{oecd2025guidelines}, \cite{oecd2024howslife}). This measure is increasingly used as a policy goal. In Britain, for example, life satisfaction is the implicit policy objective under the Treasury’s Green Book, which sets out how all policy proposals are to be appraised (\cite{hmtreasury2021wellbeing}, \cite{hmtreasury2022}).\footnote{Whether life satisfaction is the right measure of utility is contested (e.g. \citealt{benjamin_etal_2014}). We adopt it because it is a validated measure used by national statistical offices and in policy.}  


But, whether social welfare is determined by the array of incomes or life satisfaction, a key issue is the form of the social welfare function, through which individual utilities are aggregated. How egalitarian should it be? In other words, how much weight should it give to improvements for those with low scores compared with those higher up the scale?

This is an ethical question, to which there is no scientific answer. But for political decision making, it is important to know what the public thinks. That is what this paper is about.

To estimate the shape of welfare function for wellbeing, we offered a representative sample of the UK population a series of hypothetical questions. Respondents were asked to evaluate how important it is, in their opinion, to improve by one unit the life satisfaction of individuals who currently score 2 or above on a 0–10 life satisfaction scale. To establish a benchmark, the importance of improving someone’s wellbeing from 2 to 3 was fixed at 100 units. We then used the reports to estimate the shape of the welfare function without imposing parametric assumptions. We also used them to estimated the parameter of an isoelastic welfare function. 

Our analysis uses survey data collected via YouGov between 14 and 19 August 2025. The sample includes 2,068 respondents and is representative of the demographic profile of the UK adult population by age, gender, social grade, and ethnicity. We also collect respondent's own levels of life satisfaction. 

We find that the median respondent values an improvement in wellbeing from 2 to 3 as roughly twice as important as an improvement from 9 to 10. For some intermediate values, moving someone from 4 to 5 is judged about 60\% more valuable than moving someone from 8 to 9. 

With this data, we can also estimate an inequality aversion parameter $\alpha$ based on the social welfare function $(1-\alpha)^{-1} \sum_{i} U_i^{1-\alpha}k$, where $U_i$ in individual cardinal utility and $k$ is a constant. Using median values, the estimate of ${\alpha}$ is 0.48 ($\text{standard error}=0.013$). Substituting $\alpha \approx 0.5$ into the social welfare function above gives the simple approximation $S \propto \sum_{i} \sqrt{U_i}$. Aversion to wellbeing inequality not dissimilar to the aversion to income inequality estimated using hypothetical redistributive choices, i.e. withdrawing a sum of money from a higher income quantile and transferring a part of it to a lower quantile. The estimates of $\alpha$ from that approach typically lie between 0.1 and 0.5 \citep[e.g.][]{amiel_measuring_1999, pirttila_leaky_2010}. That estimate is lower than those for income found when respondents are asked to compare hypothetical societies: one with lower mean income but less inequality, versus another with higher mean income but greater inequality. These estimates range from 1 to over 3 \citep[e.g.][]{hurley_inequality_2020, pirttila_leaky_2010, carlsson_are_2005, costa-font_specific_2025}. 

The overall median estimate of 0.48 may understate the public's preferred level of inequality aversion, as it includes responses from individuals who struggled with the survey task. We find that the inequality aversion parameter is close to zero among those who reported low comprehension or high difficulty of the task, but rises substantially among the majority who found the task clear or easy.

We find significant heterogeneity in wellbeing inequality aversion across individual characteristics. Aversion to wellbeing inequality is lower among happier individuals and higher among those who engage in charitable giving. Younger respondents show higher inequality aversion than older respondents. Men have higher inequality aversion than women. Widowed respondents have markedly lower inequality aversion than other groups. Higher-income respondents display greater inequality aversion. Importantly, we find no significant differences across UK nations, implying that a single welfare weighting framework can be applied consistently for policy evaluation in the UK.

Our paper contributes to two strands of literature. First, it connects to the literature on wellbeing policy evaluation, which uses reports of life satisfaction as an overall welfare measure to assess the impact of policies or to assign value to their intangible effects.\footnote{See \citet{frijters_and_krekel_2021, frijters_et_al_2020, layard_living_2021} for formulations of this approach, which has been applied to value the benefits of volunteering programmes \citep{krekel_et_al_2024}, hosting sports events \citep{dolan_et_al_2019}, and a wide range of policies \citep{frayman_value_2025}.} Current evaluations typically treat wellbeing improvements as equally valuable, regardless of the initial wellbeing level of the recipient. Our contribution is to provide a tool for differentially valuing improvements across the wellbeing distribution, where the differences in treatment reflect the views of the UK population.

Secondly, we contribute to the literature on inequality measurement and distributional weights \citep{Atkinson1970, clark_attitudes_2015, jenkins_getting_2024, cowell_measuring_2025}.\footnote{This literature primarily uses two types of survey instruments. One approach asks respondents to evaluate hypothetical redistribution scenarios, such as withdrawing money from a higher income quantile and transferring part of it to a lower one \citep{SaezStantcheva2016, CapozzaSrinivasan2025}. Another asks respondents to compare income distributions that differ in both range and average \citep{hurley_inequality_2020, pirttila_leaky_2010}.} Our contribution here is twofold. First, we propose a novel survey instrument that {directly elicits} respondents' distributional preferences for wellbeing, rather than {recovering} them indirectly from complex hypothetical choices. This makes it easier for respondents to provide consistent answers across different points of the distribution. Second, we apply this direct-elicitation approach to {wellbeing}, which, to the best of our knowledge, has not been done before. 

\section{The Survey and Data}
\label{survey_instrument}

\begin{figure}[H]
    \centering
    \begin{subfigure}[t]{\textwidth}
        \centering
        \includegraphics[width=\textwidth]{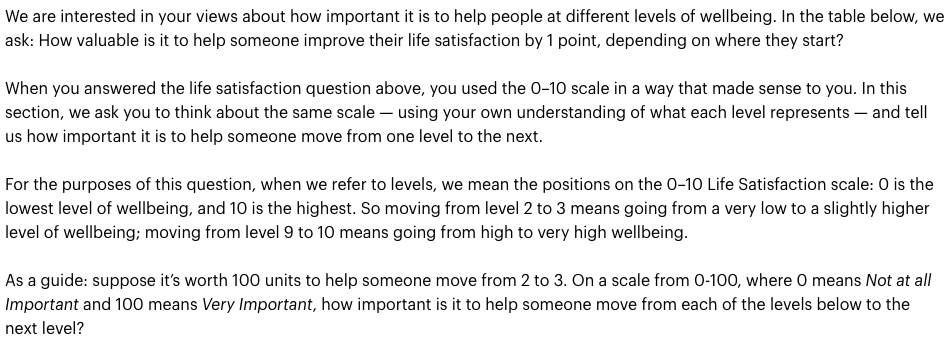}
        \label{fig:q4_part_a}
    \end{subfigure}
    
    
    \begin{subfigure}[t]{\textwidth}
        \centering
        \includegraphics[width=\textwidth]{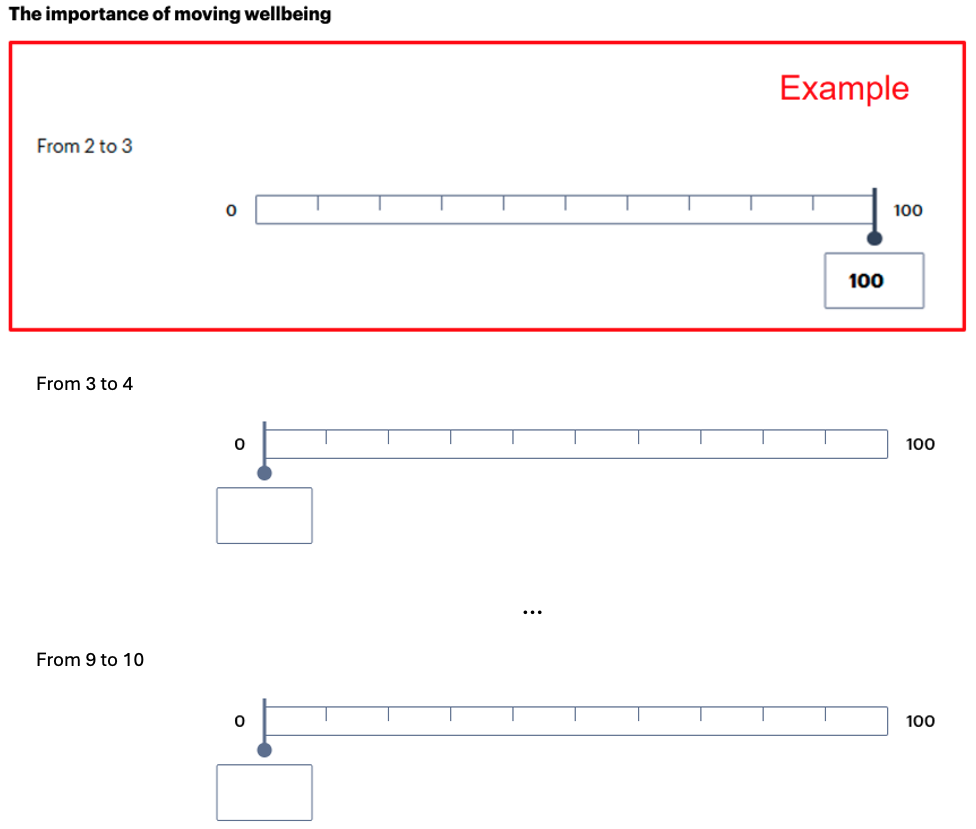}
        \label{fig:q4_part_b}
    \end{subfigure}

    \caption{Relative importance of increasing wellbeing question.}
    \label{fig:q4_main_text}
    \footnotesize
    \textit{Note:} Respondents were asked to move the dial for each life-satisfaction step between 0 and 100. The current value was shown in the box below the dial. Subsequent steps were presented further down on the same screen.
\end{figure}

Rather than infer distributional weights from choices between alternative outcomes, we elicit them by asking people directly. First, we introduce the respondents to the scale, by asking them to record their own life satisfaction (LS) on the 0-10 scale.\footnote{We asked the standard life satisfaction question about the respondent’s own life: \emph{``Overall, how satisfied are you with your life nowadays?''} (0--10 scale). } Next we tell them the distribution of actual answers in the UK population. Then we ask the key question presented in Figure \ref{fig:q4_main_text}: \emph{``How valuable is it to help someone improve their life satisfaction by 1 point, depending on where they start?''}

Proceeding step by step yields a set of relative importance weights of improving wellbeing across the whole scale from $LS = 2$ up to $LS = 9$. To make responses comparable across individuals, we fix the importance of moving from 2 to 3 at 100 units. We anchor at 2 to 3 rather than 0 to 1 because few respondents report life satisfaction below 2 in large UK and international surveys, making such states harder to imagine.  For example, the UK Office for National Statistics reports that only around 1\% of respondents reported life satisfaction below 2 \citep{ons_2023}. Starting at 2 avoids this problem while still capturing the lower end of the wellbeing distribution.

We use survey data collected via YouGov between 14 and 19 August 2025. The sample includes 2,068 respondents and is representative of the UK adult population by age, gender, social grade, and ethnicity. Apart from demographic and socio-economic characteristics, respondents report on civic engagement and other topics. We also record  respondent’s own life satisfaction score and two measures of task comprehension: whether the question made sense to them and how difficult it was to answer. See Appendix \ref{data} for the details of the sample. Appendix~\ref{app:questionnaire} further describes the survey instrument and reports screenshots of the instructions and items as presented to respondents.

\section{Estimation}
\label{section:estimation}

Before going further, it is worth noting the necessary properties of a reasonable social welfare function. First, as Arrow showed, utility must be measured in a way that is cardinal and interpersonally comparable (\cite{arrow1970}). We proceed under the assumption that life satisfaction, like income, has these characteristics. This assumption is supported by some empirical evidence. For example, test-retest errors are similar at all points on the scale \citep{krueger_reliability_2008}, suggesting that the points are equally spaced. Moreover, replies on a continuous visual analogue scale are similar to those using a discrete scale \citep{kaiser_measuring_2025}. 

As regards interpersonal comparability of utilities, this is surely not perfect, but again, without a substantial degree of comparability, we would not observe any substantial correlations of life satisfaction with objective variables and behaviours.

The final requirement of a social welfare function is symmetry. In other words, the value of the function would be unchanged if the utility of different participants were switched around. One symmetric function is the isoelastic function used by many writers, where social welfare ($S$) is given by
\begin{equation}
S = \frac{1}{1-\alpha} \sum_{i=1}^N U_{i}^{1-\alpha} k \quad (\alpha \geq 0, \alpha \neq 1, k > 0)
\end{equation}
where $U_i$ is the income or life satisfaction of person $i$ and $k$ reflects the units of $S$ and $U$ (e.g. \cite{atkinson_lectures_1980}). With this function the marginal social value of utility is
\[
\frac{\partial S}{\partial U_{i}} = U_{i}^{-\alpha} k
\]
and, taking logs of both sides: 
\begin{equation}
\log \frac{\partial S}{\partial U_{i}} = -\alpha \log U_{i} + \log k
\label{eq:2}
\end{equation}
In other words, the marginal social value of extra utility is inversely related to the level of utility with an elasticity of minus $\alpha$.

So $\alpha$ is a measure of inequality aversion. If $\alpha=0$, we have the approach proposed by Jeremy Bentham where social welfare is measured by the simple addition of utilities -- an approach also advocated by the philosopher Peter Singer (\cite{bentham1789}, \cite{singer1981}). If $\alpha \to \infty$, we have the extremely egalitarian position advocated by John Rawls where all that matters is the utility of the least happy person \cite{rawls_theory_1971}.

In this paper, we estimate equation \ref{eq:2} above. For each of the $N$ survey respondents, we have their reports on the social value of an extra unit of utility starting from each of the 7 freely chosen levels (3 up to 9). Thus, equation \ref{eq:2} is estimated using $7 \times N$ paired observations on $U_i$ and $\partial S/\partial U_{i}$.\footnote{To be explicit, there are $N$ individuals, denoted by $i$, and 7 relevant levels of utility, denoted by $j$. Each individual reports, for each of these 7 levels of $U_{ij}$, an assessment of its social marginal value $\partial S/\partial U_{ij}$. The estimation equation is thus:

\[
\log \frac{\partial S}{\partial U_{ij}} =  -\alpha \log U_{ij} + \log k+ \epsilon_{ij},
\]
\noindent where  $U_{ij}$ takes values from 3 to 9. We fit this equation with YouGov post-stratification weights and report heteroskedasticity-robust standard errors.} The dependent variable in equation \ref{eq:2} is the individual responses in logs. Its distribution is right-skewed and we estimate equation \ref{eq:2} by median regression.\footnote{Whereas OLS regression selects the regression coefficient to minimise the sum of squared errors ($\sum \epsilon_{ij}^2$), a median regression is arguably more appropriate in that it minimises the sum of absolute errors ($\sum |\epsilon_{ij}|$). This approach is in line with the proposition that the median voter's view should be decisive.} OLS estimates are reported in Appendix \ref{app:mean}.

We shall then compare our $\alpha$ with estimates of it when utility is measured by income.\footnote{Note that when a social welfare function is estimated over income, its concavity is typically attributed to the diminishing marginal utility of money, not to inequality aversion. The underlying social welfare function is then Benthamite.} We shall also see how it varies across different population groups.

\section{Results}
\label{results}

For ethical and political reasons, we focus on the median of the distributional weights at each point on the scale, rather than the mean. Table \ref{tab:figure1} presents the results. The value of going from 2 to 3 is normalised at 100. We find that the social value of someone going from 9 to 10 is only 49, just under a half the value of raising someone from 2 to 3. 

\begin{table}[h]
\centering
\caption{Median marginal value of someone going up from life satisfaction at the level shown to the next higher level of wellbeing (from 2 to 3 =100)}
\label{tab:figure1}
\begin{tabular}{lccccccccc}
 LS step & 2→3 & 3→4 & 4→5 & 5→6 & 6→7 & 7→8 & 8→9 & 9→10 \\ \hline
Median & 100 & 83 & 80 & 72 & 69 & 60 & 50 & 49 \\
\end{tabular}
\par
\begingroup
\footnotesize
\textit{Note:} The 2→3 step is fixed at 100 by design (anchor). Steps 3→4 through 9→10 are freely chosen by respondents and enter the regression in Equation \ref{eq:2}.
\endgroup
\end{table}

\begin{figure}[h]
\centering
\includegraphics[width=0.7\textwidth]{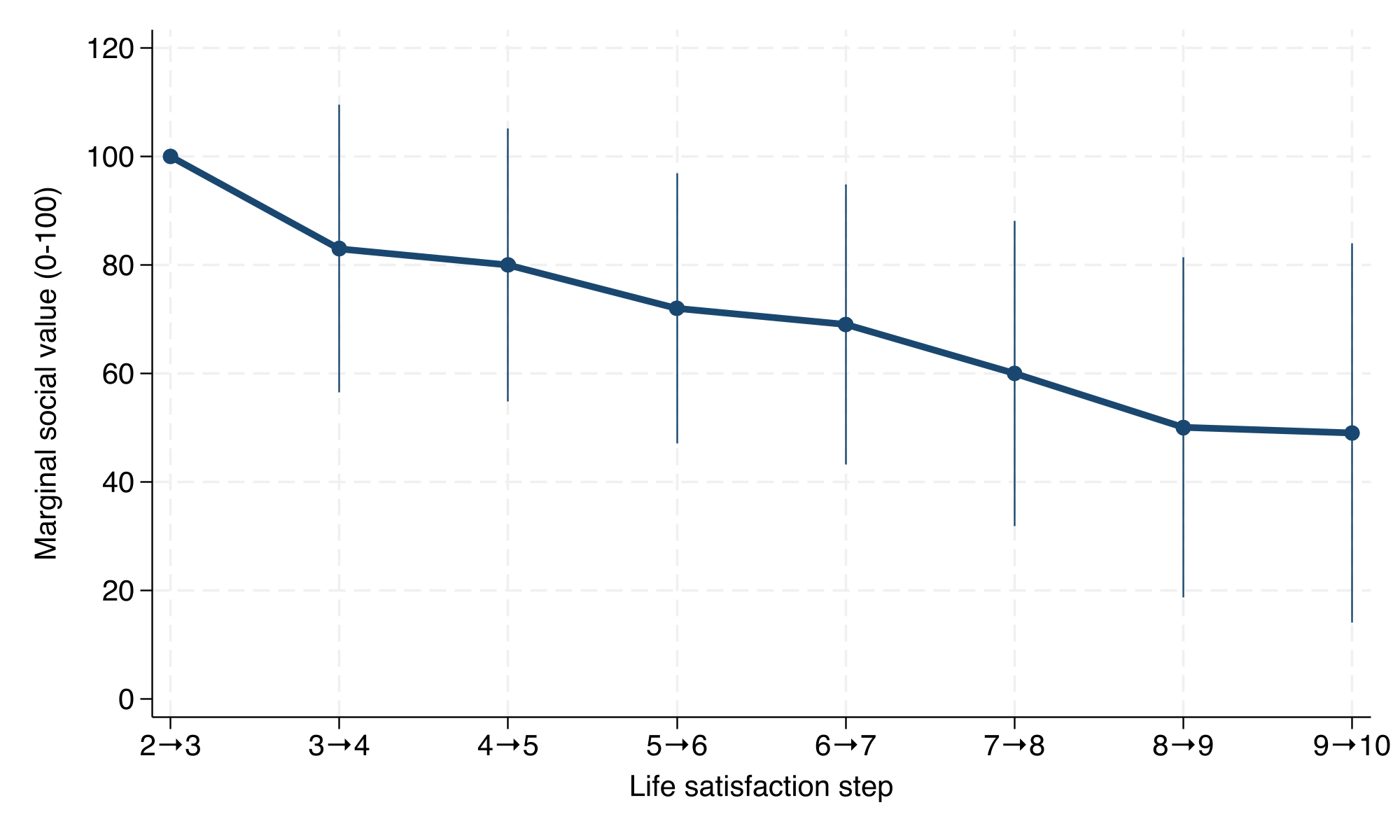}
\caption{Marginal social value when life-satisfaction is increased by one unit from different starting points (+/- 1 SD)}
\label{fig:fig_1}
\end{figure}

The same data is displayed in graphical form in Figure \ref{fig:fig_1}.\footnote{Appendix \ref{app:dist} presents the distribution of marginal social value by life satisfaction level.}

We can now estimate the value of $\alpha$, the coefficient of inequality aversion, which summarises the data in Table \ref{tab:figure1}.  We have, for each individual $i$, their estimate of all 7 freely chosen incremental values on which to run Equation \ref{eq:2}. The resulting median estimate of ${\alpha}$ is $0.48$ ($\text{se}=0.013$).\footnote{The anchor step (2→3), fixed at 100 by design, is excluded from the regression as it contains no revealed preference information. Including the anchor yields a lower median estimate of $\alpha = 0.39$ ($\text{se} = 0.020$). The corresponding OLS estimate is essentially unchanged at 0.62. For the main analysis, we drop observations where individuals reported a value of 0 due to the log-linearisation. There are 353 zero reports; recoding these as zeros rather than dropping them yields $\alpha = 0.42$ ($\text{se} = 0.014$). The mean estimate is $\alpha = 0.62$ ($\text{se} = 0.021$); full mean results are reported in Appendix \ref{app:mean}. Diagnostic checks for the quality of fit of Equation \ref{eq:2} are presented in Appendix \ref{app:diag}.}

It is interesting to compare this degree of aversion to life-satisfaction inequality with the aversion to income inequality in other studies. The latter is typically elicited in two different ways. In one respondents are asked whether they approve of a change where someone (poorer) gains $x$ and another person (richer) loses $y$. The estimates of $\alpha$ here are usually between 0.1 and 0.5 \citep[e.g.][]{amiel_measuring_1999, pirttila_leaky_2010}.  This is not that different from the coefficient on wellbeing inequality. An alternative approach asks respondents to choose between hypothetical societies with different income distributions: one with lower mean income but less inequality, versus another with higher mean income but greater inequality. Estimated using this approach the elasticity varies between 1 and over 3 \citep[e.g.][]{hurley_inequality_2020, pirttila_leaky_2010, carlsson_are_2005}. 

The closest analysis to ours is a recent work by \citet{cooper_individual_2026}. They offered 300 UK adults a series of gambles between life satisfaction levels to estimate inequality aversion. Each gamble compared a sure outcome with a lottery between a higher and a lower outcome. They find substantially higher inequality aversion than we do. To compare, we set the value of moving someone from 2 to 4 in their analysis at 100. Then moving from 8 to 10 is worth between about 2 and over 12, depending on the aggregation method.\footnote{The lower bound of 2 is calculated using $\lambda = 4.0$, the pooled median societal loss-aversion ratio for non-fatal gambles in their Table~2. Estimates kindly provided by Crispin Cooper for alternative aggregation methods across individuals yield 7.4 and 12.8, the latter incorporating a correction for the overestimation of small probabilities.} In our analysis it is above 50. This is in line with the income literature: indirect methods, lotteries or comparisons of hypothetical distributions, produce higher estimates than direct redistributive choices. \\

\noindent \textbf{Robustness of the results to question comprehension.}  A natural question about our results is ``Do people really understand the question?''. As Table \ref{tab:comprehension} shows, the survey task was cognitively demanding for some respondents. And this affected their answers. People who found the question difficult reported much lower inequality aversion than people who found it easy (see Table \ref{tab:alpha_combined}).\footnote{$\alpha$ estimates from Equation \ref{eq:2} with interactions for comprehension levels. Joint F-tests examine equality across comprehension response categories.} This suggests that the overall estimate of 0.48 may understate the public's level of inequality aversion, as it includes responses from individuals who struggled with the survey task.

\begin{table}[htbp]
\centering
\caption{Question comprehension and difficulty}
\label{tab:comprehension}
\begin{tabular}{lclc}
\multicolumn{2}{l}{Did the questions make sense to you?} & \multicolumn{2}{l}{How easy or difficult it was to answer?} \\
Response & Percent & Response & Percent \\
\midrule
1 (Not at all) & 21 & 1 (Very difficult) & 16 \\
2 & 13 & 2 & 17 \\
3 & 20 & 3 & 27 \\
4 & 19 & 4 & 19 \\
5 (Completely) & 27 & 5 (Very easy) & 21 \\
\bottomrule
\end{tabular}
\end{table}

\begin{table}[htbp]
\centering
\caption{Inequality aversion by question comprehension}
\label{tab:alpha_combined}
\begin{tabular}{lcclcc}
\multicolumn{3}{c}{Did the questions make sense to you?} & \multicolumn{3}{c}{How easy or difficult it was to answer?} \\
Response & $\alpha$ & se & Response & $\alpha$ & se \\
\midrule
Not at all & $-$0.10 & 0.035 & Very difficult & $-$0.09 & 0.040 \\
2 & 0.36 & 0.038 & 2 & 0.39 & 0.027 \\
3 & 0.43 & 0.028 & 3 & 0.55 & 0.020 \\
4 & 0.68 & 0.024 & 4 & 0.64 & 0.027 \\
Completely & 0.61 & 0.056 & Very easy & 0.51 & 0.068 \\
\midrule
\multicolumn{6}{c}{Joint test \textit{p-value}: Clarity $< 0.001$, Difficulty $< 0.001$}
\end{tabular}
\par
\begingroup
\footnotesize \singlespace 
\textit{Notes:} $\alpha$ estimated by median regression from Equation \ref{eq:2} with interactions for question comprehension and difficulty. Joint F-test examines equality of $\alpha$ across all levels.
\endgroup
\end{table}

\begin{figure}[htbp]
\centering
\begin{subfigure}{0.4\textwidth}
    \includegraphics[width=\linewidth]{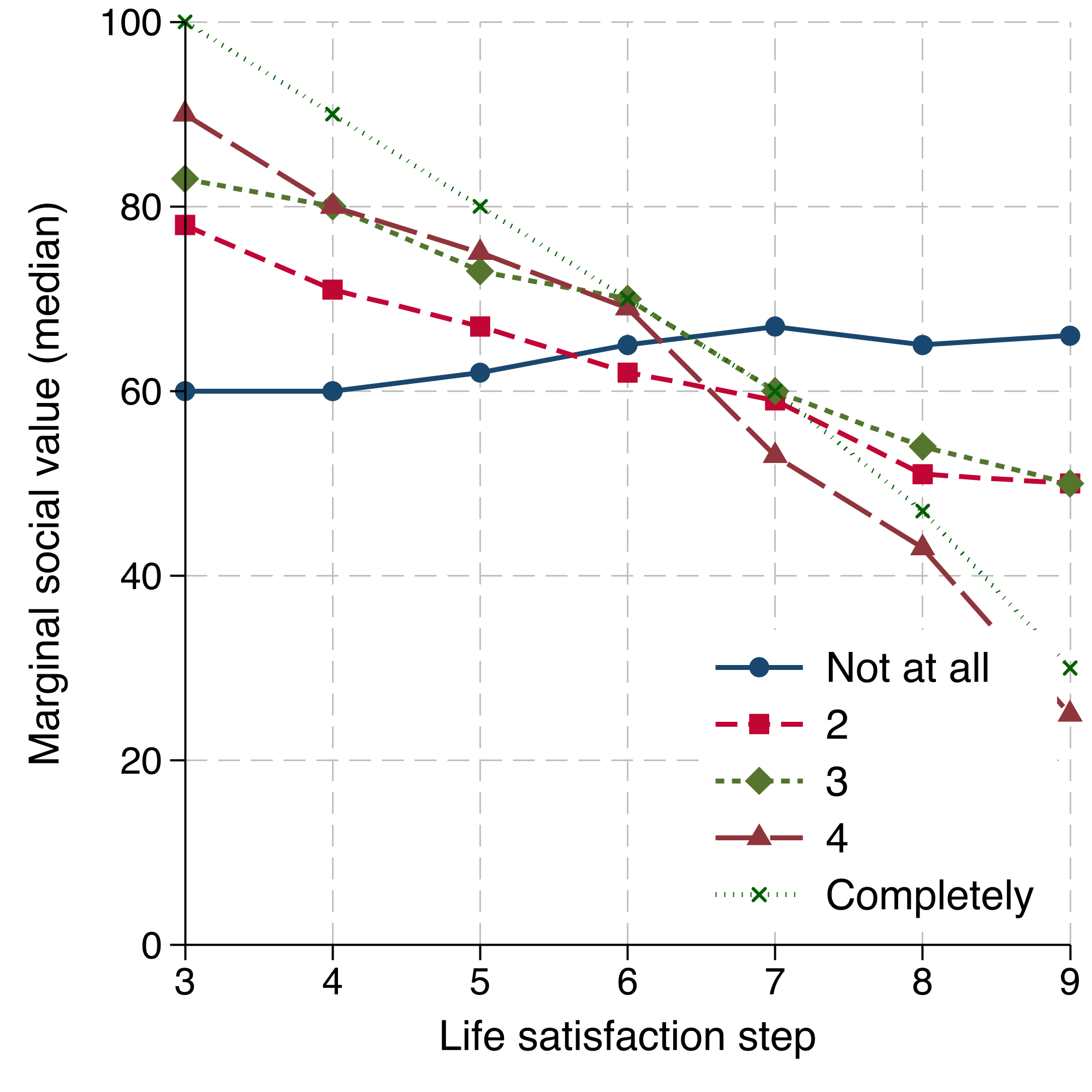}
    \caption{By question clarity}
\end{subfigure}
\begin{subfigure}{0.4\textwidth}
    \includegraphics[width=\linewidth]{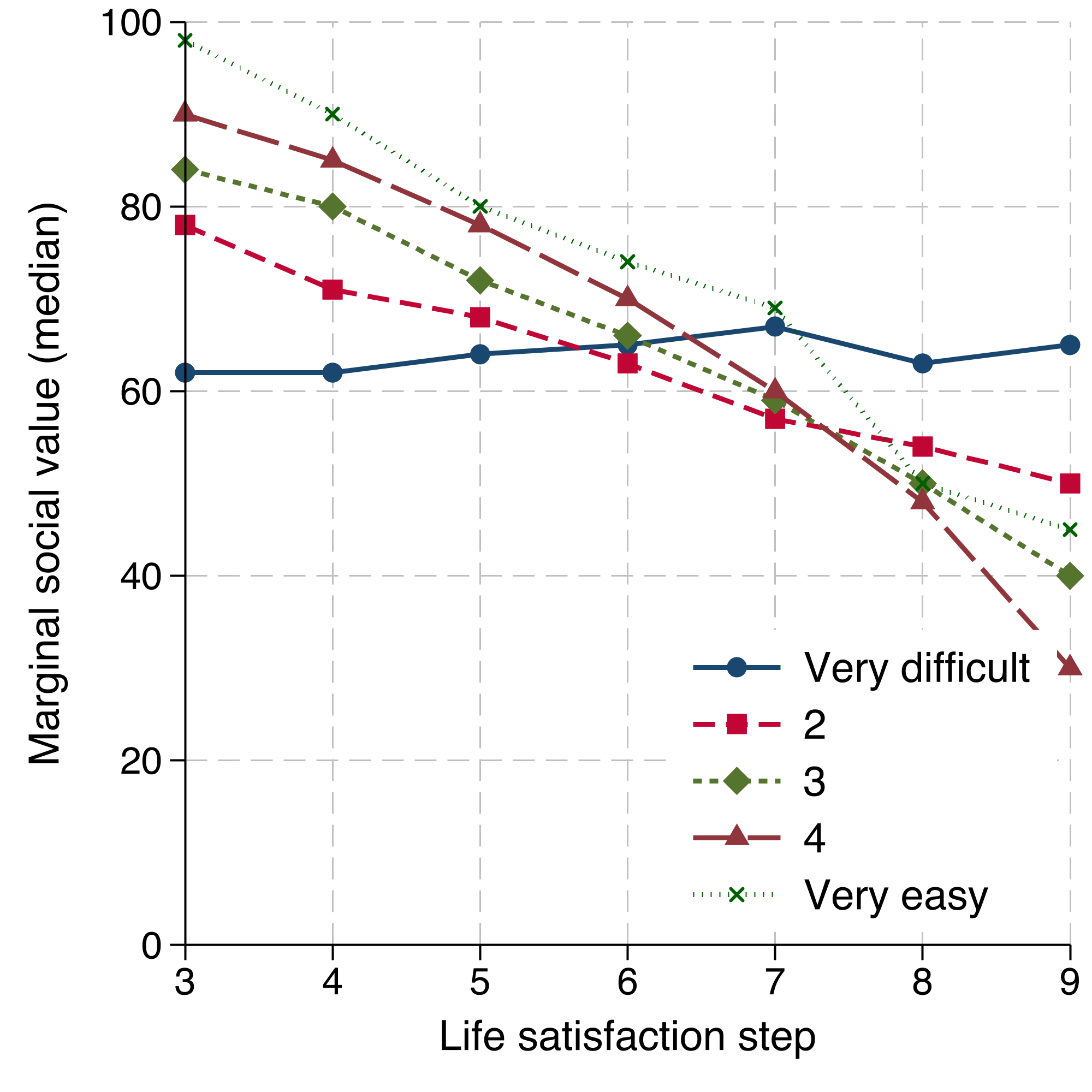}
    \caption{By question difficulty}
\end{subfigure}
\caption{Marginal social value by survey evaluation questions}
\label{fig:fig_clear_easy}
\end{figure}

For example, 21\% of respondents reported that the questions did not make sense at all, and 16\% found them very difficult. $\alpha$ equals $-0.10$ for those who found that the questions did not make sense at all as against 0.61 for those who found that they made complete sense. Similarly, $\alpha$ equals $-0.09$ for those who found the questions very difficult and 0.51 for those who found them very easy. Figure \ref{fig:fig_clear_easy} shows that respondents who struggled with the question tended to report the same marginal social value for all levels of life satisfaction. 

To better understand whether the clarity and difficulty of the task are systematically related to respondent characteristics, we regress clarity and easiness responses on a set of individual controls (see Appendix \ref{app:rob}). The main correlate that we identify is the education level: respondents with higher education levels reported that the questions were clearer and easier to answer. Those who engaged in charitable giving found the questions more clear, but not easier to answer.\footnote{Column 3 of Table \ref{tab:regressions} regresses life satisfaction on the same set of controls. The relationships are consistent with the wellbeing literature.}\\

\noindent \textbf{Heterogeneity.} We shall now examine how inequality aversion differs for different groups. First, people who are themselves happy are less averse to wellbeing inequality than those less fortunate (see Table \ref{tab:alpha_lsat}). This echoes the findings for inomes of \cite{beckman_experimental_2006}, \cite{amiel_measuring_1999} and \cite{CapozzaSrinivasan2025} that richer respondents are less averse to income inequality than poor ones are.

\begin{table}[htbp]
\centering
\caption{Inequality aversion by individual's life satisfaction}
\label{tab:alpha_lsat}
\begin{tabular}{lcc}
Life Satisfaction & $\alpha$ & se \\
\midrule
$\leq 2$ & 0.53 & 0.072 \\
3 & 0.57 & 0.046 \\
4 & 0.57 & 0.057 \\
5 & 0.48 & 0.043 \\
6 & 0.52 & 0.031 \\
7 & 0.53 & 0.025 \\
8 & 0.32 & 0.023 \\
9 & 0.24 & 0.061 \\
10 & 0.14 & 0.061 \\
\midrule
\multicolumn{3}{l}{Joint test of equality: \textit{p-value} $< 0.001$}
\end{tabular}
\par
\begingroup
\footnotesize
\textit{Notes:} $\alpha$ estimated by median regression from Equation \ref{eq:2} with interactions for individual's own life satisfaction levels. Life satisfaction levels 0, 1 and 2 are merged into a single bottom category due to small cell sizes. Joint F-test examines equality of $\alpha$ across all satisfaction levels.
\endgroup
\end{table}

People engaged in financial giving are more inequality averse than others ($\alpha = 0.49$ \textit{vs.} $0.36$), suggesting that individual behaviours are informative about their preferred shape of welfare function. We find no significant differences by volunteering or health status (see Table \ref{tab:alpha_group2}).

\begin{table}[htbp]
\centering
\caption{Inequality aversion by charitable giving and health}
\label{tab:alpha_group2}
\begin{tabular}{lclclc}
\multicolumn{2}{c}{Financial giving} & \multicolumn{2}{c}{Volunteering} & \multicolumn{2}{c}{Health} \\
Response & $\alpha$ (se) & Response & $\alpha$ (se) & Response & $\alpha$ (se) \\
\midrule
No & 0.36 (0.024) & No & 0.48 (0.015) & Limited & 0.48 (0.025) \\
Yes & 0.49 (0.016) & Yes & 0.50 (0.040) & Not limited & 0.48 (0.015) \\
\midrule
\multicolumn{6}{l}{Joint test \textit{p-values}: Financial giving $< 0.001$, Volunteering = 0.613, Health = 1.000}
\end{tabular}
\par
\begingroup
\footnotesize
\textit{Notes:} $\alpha$ estimated by median regression from Equation \ref{eq:2} with interactions for the levels of each variable. Joint F-test examines equality of $\alpha$ across all levels of each variable. 
\endgroup
\end{table}

Younger respondents show significantly higher inequality aversion ($\alpha = 0.55$ for 18--34, compared to $0.36$ for 55+). This is consistent with earlier results for income-related inequality, where younger respondents also have higher inequality aversion \citep[see e.g.][]{CapozzaSrinivasan2025}. Men have significantly higher inequality aversion than women (0.49 \textit{vs.} 0.38). Widowed respondents have markedly lower inequality aversion (0.17) than married/partnered (0.48) or divorced/separated (0.51) respondents (see Table \ref{tab:alpha_group3}).

\begin{table}[htbp]
\centering
\caption{Inequality aversion by demographic characteristics}
\label{tab:alpha_group3}
\begin{tabular}{lccccc}
\multicolumn{2}{c}{Age group} & \multicolumn{2}{c}{Gender} & \multicolumn{2}{c}{Marital status} \\
\cmidrule(lr){1-2}\cmidrule(lr){3-4}\cmidrule(lr){5-6}
Response & $\alpha$ (se) & Response & $\alpha$ (se) & Response & $\alpha$ (se) \\
\midrule
18-34 & 0.55 (0.024) & Male & 0.49 (0.023) & Married / Partnered & 0.48 (0.017) \\
35-54 & 0.46 (0.022) & Female & 0.38 (0.015) & Divorced / Separated & 0.51 (0.028) \\
55+ & 0.36 (0.025) & & & Single & 0.39 (0.056) \\
 & & & & Widowed & 0.17 (0.065) \\
\midrule
\multicolumn{6}{l}{Joint test \textit{p-values}: Age group $< 0.001$, Gender $< 0.001$, Marital status $< 0.001$}
\end{tabular}
\par
\begingroup
\footnotesize
\textit{Notes:} $\alpha$ estimated by median regression from Equation \ref{eq:2} with interactions for the levels of each variable. Joint F-test examines equality of $\alpha$ across all levels of each variable.
\endgroup
\end{table}

Higher-income respondents display greater inequality aversion: $\alpha = 0.54$ for those earning \pounds60,000 or more per year, compared with $\alpha = 0.44$ for those earning below \pounds20,000 (joint $p$-value $= 0.034$). We find no significant differences across UK nations (see Table \ref{tab:alpha_income_country}). This is reassuring for policy evaluation, as it suggests that a single welfare weighting framework can be applied consistently across the whole of the UK without region-specific adjustments.

\begin{table}[htbp]
\centering
\caption{Inequality aversion by household income and country}
\label{tab:alpha_income_country}
\begin{tabular}{lccc}
\multicolumn{2}{c}{Household income} & \multicolumn{2}{c}{Country} \\
\cmidrule(lr){1-2}\cmidrule(lr){3-4}
Response & $\alpha$ (se) & Response & $\alpha$ (se) \\
\midrule
Up to \pounds19,999 & 0.44 (0.037) & England & 0.48 (0.015) \\
\pounds20,000--39,999 & 0.46 (0.033) & Wales & 0.52 (0.064) \\
\pounds40,000--59,999 & 0.49 (0.021) & Scotland & 0.41 (0.035) \\
\pounds60,000+ & 0.54 (0.020) & Northern Ireland & 0.48 (0.061) \\
\midrule
\multicolumn{4}{l}{Joint test \textit{p-values}: Income = 0.034, Country = 0.247}
\end{tabular}
\par
\begingroup
\footnotesize
\textit{Notes:} $\alpha$ estimated by median regression from Equation \ref{eq:2} with interactions for the levels of each variable. Joint F-test examines equality of $\alpha$ across all levels of each variable.
\endgroup
\end{table}

\section{Illustrative policy example}

Finally, we shall now illustrate how using distributional weights can change the rankings between policies. Suppose two policies both cost the same. But

\begin{itemize}
    \item Policy A moves 1,000 people from level 2 to 3
    \item Policy B moves 1,500 people from level 8 to 9
\end{itemize}

\noindent Without distributional weights, Policy B appears better (helping $1,500 > 1,000$ people). But if we add distributional weights, Policy A is best since

\begin{itemize}
    \item {Policy A welfare benefit:} $1,000 \times 100 = 100,000$ 
    \item {Policy B welfare benefit:} $1,500 \times 50 = 75,000$,
\end{itemize}

\noindent where 100 and 50 are the respective median marginal welfare weights for improvements at levels 2 and 8.

\section{Conclusion}

We have estimated the form of the UK public's social welfare function, defined across individual cardinal utilities. In a survey of 2,068 adults we asked them to state the importance of an extra unit of utility starting from each level (with the transition from 2 to 3 being valued at 100). The results are shown in Figure \ref{fig:fig_1}.

It is convenient to summarise these numbers by assuming they emerge from a function where social welfare ($S$) is given by
\[
S = \frac{1}{1-\alpha} \sum_{i=1}^n U_{i}^{1-\alpha} k \quad (\alpha \geq 0, \alpha \neq 1, k > 0)
\]
where $U_{i}$ is the utility of the person $i$. Differentiating by $U_i$ makes the marginal social utilities proportional to $U_{i}^{-\alpha}$. When we estimate this function using median regression we find that $\alpha$ is equal to 0.48 ($se = 0.013$). 

Our findings provide ethically grounded distributional weights for wellbeing policy evaluation and cost-benefit analysis. We demonstrate that incorporating these weights can alter the ranking of policy options, ensuring public resources are allocated in a way that reflects the values of the population. 



\bibliographystyle{apalike}  
\bibliography{bibl}

\clearpage
\appendix

\section{Data and Sample}
\label{data}

\subsection{Data Collection}

Fieldwork was conducted by YouGov between 14 and 19 August 2025. The study is based on an online self-completion survey administered to members of the YouGov UK panel, which consists of more than 2.5 million individuals who have agreed to participate in research. For this study, email invitations were issued at random from the base sample. Invitations contained a generic survey link. Once a panel member clicked on the link, they were directed to the study questionnaire. 

The project was reviewed in line with the {LSE Research Ethics Policy and Procedure}. Prior to the survey with YouGov, the instrument was piloted with a group of postgraduate students at LSE. All respondents included in the analysis provided informed consent before beginning the survey.

\subsection{Sample}

The full sample includes 2,111 observations. Of these, 43 respondents did not consent to continue the survey, leaving a final sample of 2,068 individuals.

Table \ref{table:desc_cov} reports descriptive statistics for the sample. Around 52\% of respondents are female, 27\% are aged between 18 and 34, 33\% are aged between 35 and 54, and 40\% are 55 or older. 86\% of the sample identify as white, 58\% are married or partnered, and 53\% report having no religious faith. 60\% are employed full-time or part-time, and 39\% work in the private sector. In terms of civic engagement, 47\% report having given money to charity and 11\% report volunteering in the past 12 months.

Table \ref{table:desc_outcomes} presents descriptive statistics for life satisfaction responses and follow-up questions on clarity and difficulty. 43\% of respondents report life satisfaction of 7 or 8 on the 0–10 scale, 10\% score 9 or 10, and 20\% report a score of 4 or below. For question comprehension, around 65\% of respondents reported that the question made some sense (3 or above on the 1--5 scale, where 1 is \textit{not at all clear} and 5 is \textit{completely clear}), and around 67\% found the question somewhat easy to answer (3 or above on the 1--5 scale, where 1 is \textit{very difficult} and 5 is \textit{very easy}). However, 21\% of respondents reported that the question was \textit{not at all clear}, the lowest category, and 16\% described it as \textit{very difficult}, the lowest category.

All descriptive statistics and analyses in the paper use YouGov’s post-stratification weights to ensure national representativeness. These weights adjust the achieved sample to match the demographic profile of the UK adult population on age, gender, region, and social grade.

\subsection{Descriptive Statistics}

{\singlespacing
\begin{longtable}{l*{1}{cc}}
\caption{Distribution of life satisfaction etc. across individuals.}
\label{table:desc_outcomes} \\
&  Mean & SD\\
\hline
\endhead
\hline
Observations  &  2068&      \\
\endfoot
\textit{Life Satisfaction:}&  &      \\
Not at all    &  0.03&  0.17\\
1 &  0.01&  0.10\\
2 &  0.04&  0.19\\
3 &  0.06&  0.25\\
4 &  0.06&  0.24\\
5 &  0.12&  0.33\\
6 &  0.14&  0.35\\
7 &  0.23&  0.42\\
8 &  0.20&  0.40\\
9 &  0.06&  0.24\\
Completely    &  0.04&  0.20\\
&&\\
\textit{Did the questions make sense:}&  &      \\
Not at all    &  0.21&  0.41\\
2 &  0.13&  0.33\\
3 &  0.19&  0.40\\
4 &  0.19&  0.39\\
Completely    &  0.27&  0.4\\
&&\\
\textit{How easy was to answer:}&  &      \\
Very difficult      &  0.16&  0.37\\
2 &  0.17&  0.38\\
3 &  0.27&  0.44\\
4 &  0.19&  0.39\\
Very easy     &  0.21&  0.41\\
\hline
\end{longtable}
}

\clearpage

{\singlespacing
\begin{longtable}{l*{1}{cc}}
\caption{Descriptive statistics for the covariates.}
\label{table:desc_cov} \\
&  Mean& SD\\
\hline
\endfirsthead

\caption[]{Descriptive statistics for the covariates (cont.)}
\label{table:desc_cov} \\
&  Mean& SD\\
\hline
\endhead

\multicolumn{3}{c}{\textit{Continued on next page}} \\
\endfoot

\hline
Observations  &  2068&      \\
\endlastfoot

\textit{Gender:}    &  &      \\
Male    &  0.48&  0.50\\
Female  &  0.52&  0.50\\
&&\\
\textit{Age group:} &  &      \\
18-34   &  0.27&  0.44\\
35-54   &  0.33&  0.47\\
55+     &  0.40&  0.49\\
&&\\
\textit{Ethnicity:} &  &      \\
White   &  0.86&  0.35\\
Ethnic minority     &  0.14&  0.35\\
&&\\
\textit{Marital status:}&  &      \\
Married / Partnered &  0.58&  0.49\\
Divorced / Separated&  0.27&  0.44\\
Single  &  0.07&  0.26\\
Widowed &  0.04&  0.20\\
Missing &  0.03&  0.18\\
&&\\
\textit{Religion:}  &  &      \\
Christian     &  0.30&  0.46\\
Other faiths  &  0.09&  0.29\\
No faith      &  0.53&  0.50\\
Missing &  0.08&  0.27\\
&&\\
\textit{Sexuality:} &  &      \\
Heterosexual / Missing   &  0.88&  0.32\\
Non-heterosexual    &  0.12&  0.32\\
&&\\
\textit{Work status:}&  &      \\
Working (FT or PT)  &  0.60&  0.49\\
Unemployed / Non-working&  0.10&  0.30\\
Retired &  0.23&  0.42\\
Student / Other     &  0.07&  0.26\\
&&\\
\textit{Household income:}&  &      \\
Up to £19,999 per year&  0.16&  0.37\\
£20,000 to £39,999 per year&  0.24&  0.43\\
£40,000 to £59,999 per year&  0.15&  0.36\\
£60,000 per year or more&  0.22&  0.41\\
Missing &  0.23&  0.42\\
&&\\
\textit{Sector:}    &  &      \\
Private sector      &  0.39&  0.49\\
Public sector &  0.21&  0.41\\
Third/voluntary sector&  0.06&  0.23\\
Missing &  0.34&  0.47\\
&&\\
\textit{Household size:}&  &      \\
1 &  0.23&  0.42\\
2 &  0.36&  0.48\\
3 &  0.17&  0.38\\
4 or more     &  0.22&  0.41\\
Missing &  0.02&  0.14\\
&&\\
\textit{IMD decile:}&  &      \\
Decile 1 to 3 (most deprived)&  0.26&  0.44\\
Decile 4 to 7 (middle)&  0.42&  0.49\\
Decile 8 to 10 (least deprived)&  0.32&  0.47\\
&&\\
\textit{Education:} &  &      \\
Low     &  0.20&  0.40\\
Medium  &  0.36&  0.48\\
High    &  0.44&  0.50\\
&&\\
\textit{Media source:}&  &      \\
A printed copy of a newspaper&  0.03&  0.17\\
A newspaper's website&  0.10&  0.30\\
A news website not associated with a newspaper&  0.09&  0.29\\
A news app on a mobile or tablet device&  0.17&  0.38\\
Social network websites&  0.16&  0.37\\
Television    &  0.26&  0.44\\
Radio   &  0.09&  0.29\\
Missing &  0.09&  0.28\\
&&\\
\textit{Social grade:}&  &      \\
AB      &  0.23&  0.42\\
C1      &  0.31&  0.46\\
C2      &  0.21&  0.41\\
DE      &  0.25&  0.43\\
&&\\
\textit{Region:}    &  &      \\
England &  0.84&  0.37\\
Wales   &  0.05&  0.21\\
Scotland      &  0.08&  0.28\\
Northern Ireland    &  0.03&  0.16\\
&&\\
\textit{Disability:}&  &      \\
Limited at all      &  0.31&  0.46\\
Not limited   &  0.66&  0.47\\
Missing &  0.02&  0.14\\
&&\\
\textit{Tenure:}    &  &      \\
Own     &  0.60&  0.49\\
Rent    &  0.24&  0.43\\
Other / Neither     &  0.13&  0.34\\
Missing &  0.03&  0.17\\
&&\\
\textit{Financial giving:}&  &      \\
No      &  0.50&  0.50\\
Yes     &  0.47&  0.50\\
Missing &  0.02&  0.15\\
&&\\
Volunteered in last 12 months&  0.11&  0.31\\
\end{longtable}
}
\noindent { \footnotesize \textit{Note:} Social Grade is a classification system based on occupation used in the UK: AB - Higher and intermediate managerial, administrative, professional occupations; C1 - Supervisory, clerical, junior managerial, administrative, professional occupations; C2 - Skilled manual occupations; DE - Semi-skilled and unskilled manual occupations, unemployed and lowest grade occupations. Education levels: Low (no formal qualifications to GCSE level), Medium (A-Levels/vocational/professional qualifications), High (university degrees).}

\clearpage

\section{Questionnaire Details}\label{app:questionnaire}

The core task of the survey experiment asks respondents to judge how important it is to raise someone’s wellbeing at different points on the life satisfaction scale. This is a cognitively demanding task, so the instrument was structured to make the question as intuitive as possible while preserving the information needed for estimation. 

After informed consent, we asked the standard life satisfaction question about the respondent’s own life: \emph{``Overall, how satisfied are you with your life nowadays?''} (0--10 scale). This familiarised respondents with the response scale and ensured that, when later asked to evaluate improvements for others, they would be using the \emph{same} evaluative frame they had just applied to themselves.

Before the importance questions, participants were given brief context about how the life satisfaction scale is used in representative UK surveys, including a short description of typical response patterns. They were explicitly instructed to interpret the 0--10 scale in the same way they had just done for themselves. This framing reduces ambiguity about what counts as a ``one-point'' improvement and anchors the task in a familiar metric.

Respondents were encouraged to think about the task in a policy context, where public spending can be directed toward both those who are struggling and those who are already doing well, but resources are limited, hence, policies that support some groups of individuals need to be prioritised over others. 

We fixed a common benchmark by defining the improvement from life satisfaction from 2 to 3 to have an importance of 100 units. Respondents were then asked, for each subsequent step \(LS\to LS+1\) with \(LS=3,\dots,9\):
\begin{quote}
\small
\emph{``On a scale from 0-100, where 0 means not at all important and 100 means very important, how important is it to help someone move from each of the levels below to the next level?''}
\end{quote}
Proceeding step by step yielded a set of relative importance {weights} across the whole scale from $LS = 2$ up to $LS = 9$. 

Following best practice in survey methodology (see \citet[][Ch.~2]{tourangeau2000psychology} and \citet[Ch.~2 and Ch.~4]{willis2005cognitive}), we included post-survey debriefing items to capture respondents’ perceptions of clarity (\textit{Did the question make sense to you?}) and difficulty (\textit{How easy or difficult was it to answer the question?}) of the main question. These measures help assess clarity and the cognitive burden of the instrument and support robustness checks, where we report how the estimates of welfare function depend on perceived clarity and difficulty.

\begin{figure}[h]
    \centering
    \includegraphics[width=\textwidth]{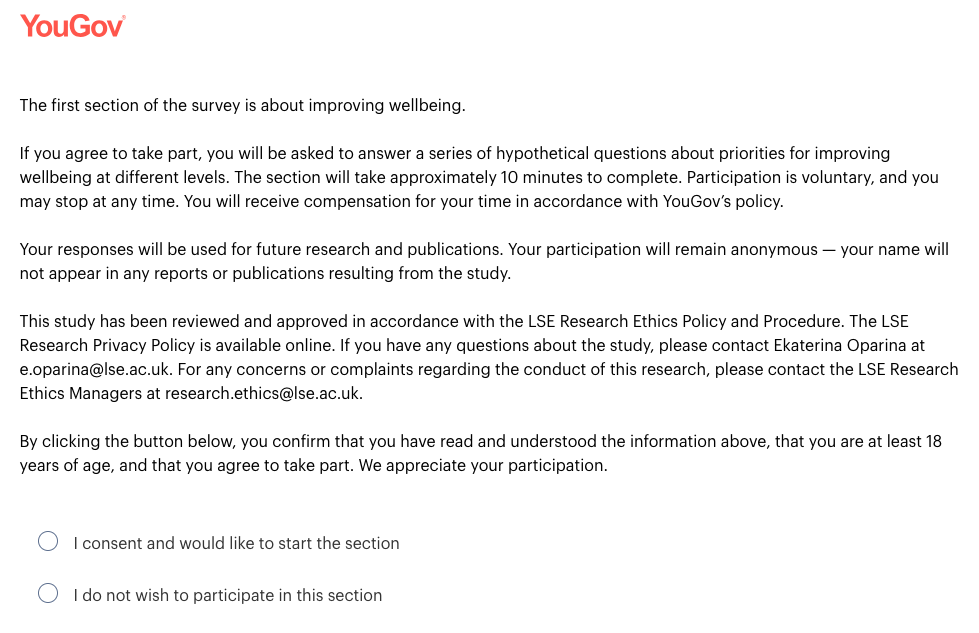}
    \caption{Consent form.}
    \label{fig:q1}
\end{figure}

\begin{figure}[h]
    \centering
    \includegraphics[width=\textwidth]{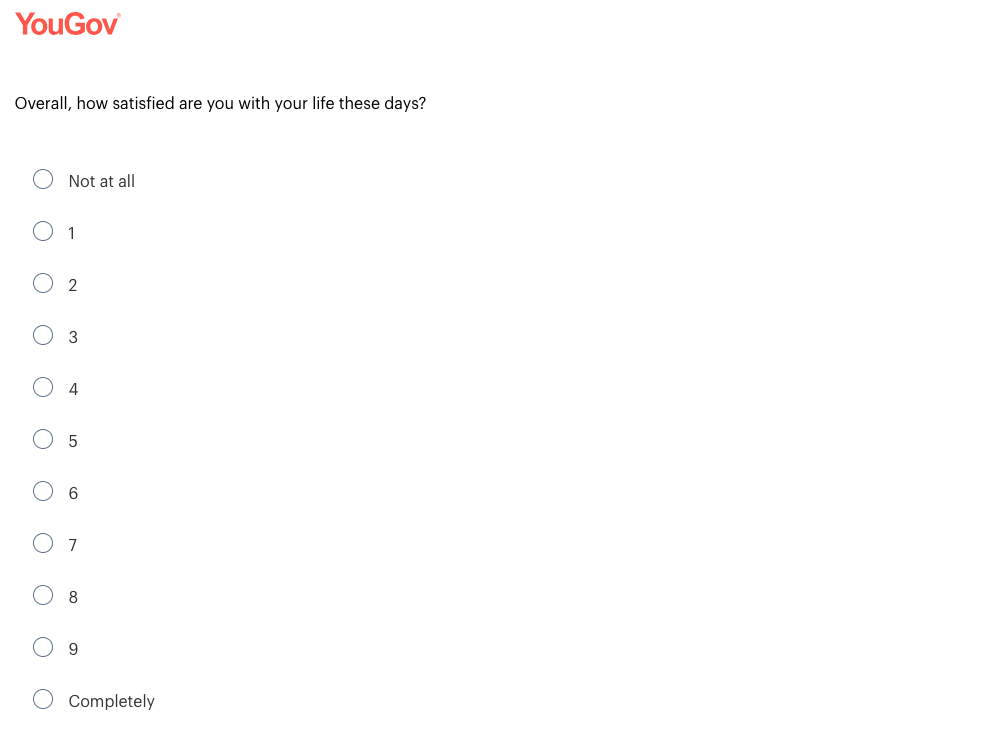}
    \caption{Life satisfaction (0--10 scale).}
    \label{fig:q2}
\end{figure}

\begin{figure}[h]
    \centering
    \includegraphics[width=\textwidth]{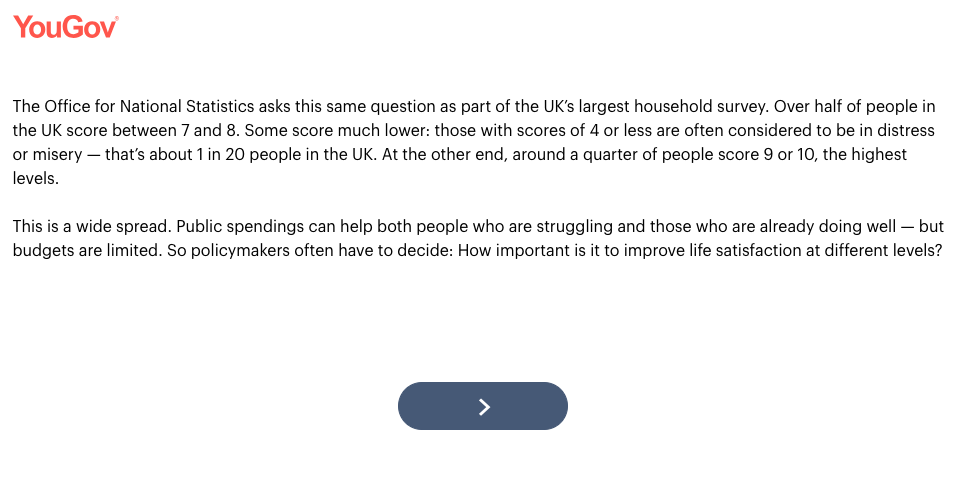}
    \caption{Relative importance of increasing wellbeing (preamble).}
    \label{fig:q3}
\end{figure}

\begin{figure}[h]
    \centering
    \includegraphics[width=\textwidth]{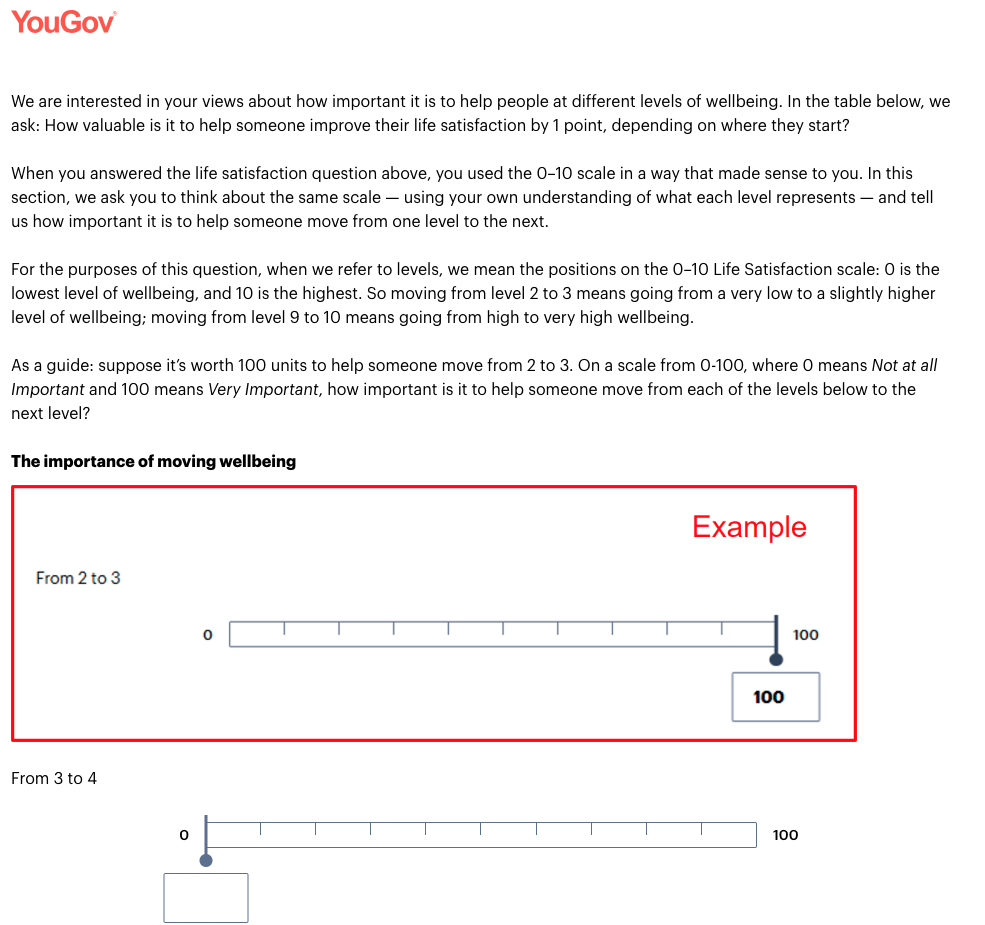}
    \caption{Relative importance of increasing wellbeing question.}
    \label{fig:q4}
\end{figure}

\begin{figure}[h]
    \centering
    \includegraphics[width=\textwidth]{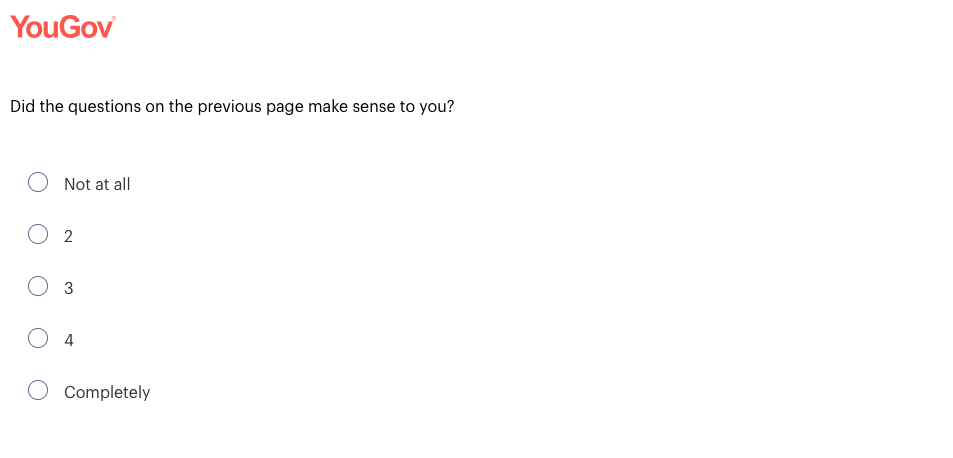}
    \caption{Follow-up comprehension question.}
    \label{fig:q5}
\end{figure}

\begin{figure}[h]
    \centering
    \includegraphics[width=\textwidth]{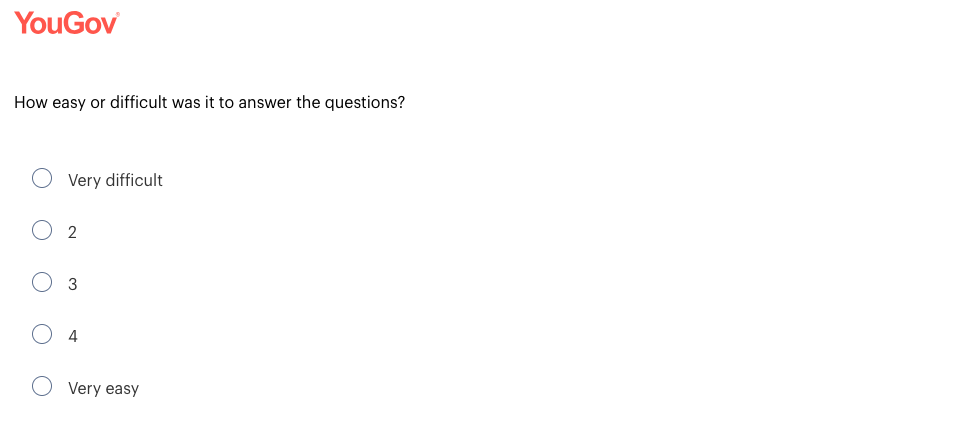}
    \caption{Follow-up difficulty question.}
    \label{fig:q6}
\end{figure}

Figures~\ref{fig:q1}--\ref{fig:q6} reproduce the survey questions as displayed to participants.

\clearpage

\clearpage

\clearpage

\section{Distribution of Marginal Social Value by Life Satisfaction Level}
\label{app:dist}

\begin{figure}[h]
    \centering
    \includegraphics[width=0.9\textwidth]{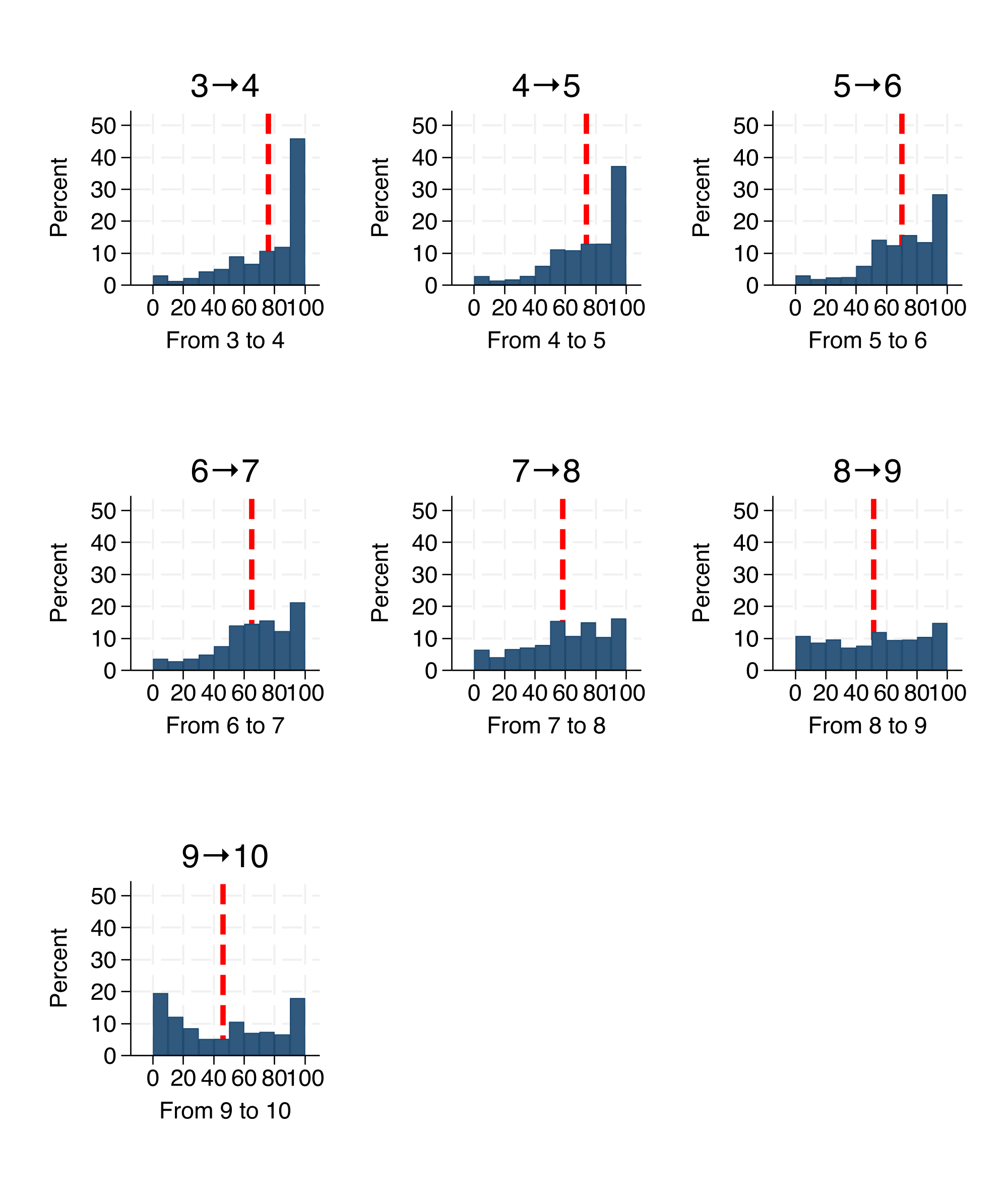}
    \caption{Distribution of marginal social value by life satisfaction level.}
    \label{fig:app_dist}
\end{figure}

\textit{Note:} Each histogram shows the distribution of marginal social value (0-100) assigned to improving life satisfaction by one point. Red line indicates mean.

\clearpage
\section{Diagnostic Checks for Equation \ref{eq:2}}
\label{app:diag}

\begin{figure}[h]
\centering
\includegraphics[width=0.9\textwidth]{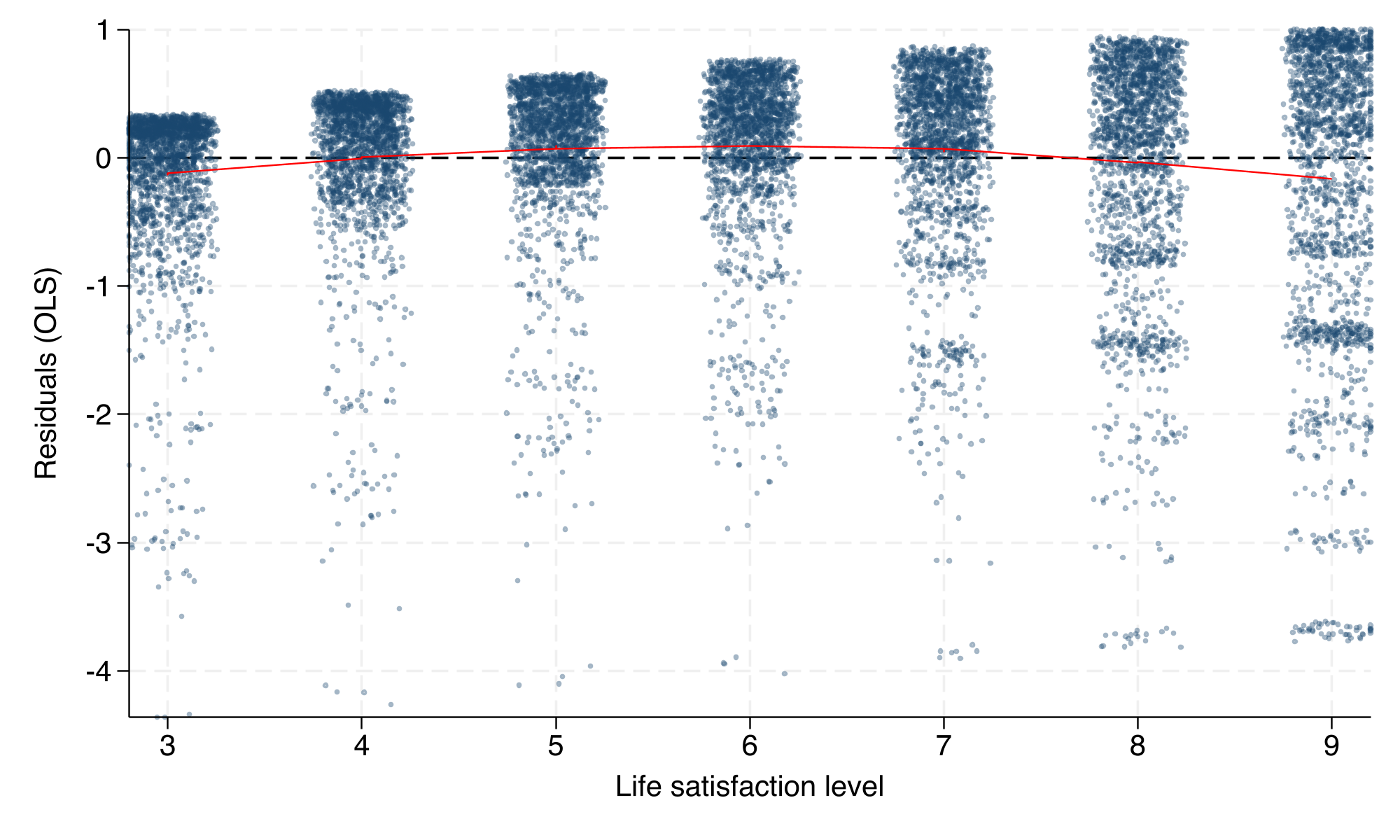}
\caption{Residuals from the OLS regression of Equation \ref{eq:2}}
\label{fig:fig_resid}
\end{figure}

To check whether the log-linear specification of the isoelastic weighting function provides a good fit, we examine the distribution of residuals from the OLS regression of the log relative weights on the log of wellbeing levels. Overall, the residual pattern supports the adequacy of the log-linear isoelastic form. Figure~\ref{fig:fig_resid} plots these residuals against the corresponding life satisfaction levels, $LS$. Each dot represents an individual residual, while the red line shows a locally weighted (Lowess) smooth of the residuals across $LS$.

The residuals are approximately centred around zero over the full range of wellbeing, indicating that the model’s predictions are, on average, unbiased. The Lowess curve is nearly flat, with only a slight downward bend at higher satisfaction levels, suggesting that the model may marginally overpredict relative weights for individuals at the top of the wellbeing distribution. The vertical spread of residuals widens modestly at higher $LS$, which points to mild heteroskedasticity. 

While most residuals are close to zero, their distribution is not symmetric: there is a concentration of positive residuals and a longer negative tail. This asymmetry motivates the use of median regression, which is robust to such skewness. The corresponding mean estimate is $\alpha = 0.62$ ($\text{se} = 0.021$); full mean results are reported in Appendix \ref{app:mean}.

\clearpage
\newpage

\section{Robustness to Question Comprehension}
\label{app:rob}

Columns 1 and 2 report the results of regression of responses about clarity (\textit{Did the question make sense to you?} from 1 -- Not at all to 5 -- Completely) and difficulty (\textit{How easy or difficult was it to answer the question?} from 1 - Very difficult to 5 -- Very easy) of the survey instrument on the set of individual controls. Column 3 reports the results of regressing individual level of life satisfaction on the same set of controls. 

{\singlespacing\small
\begin{longtable}{l*{3}{c}}
\caption{Regression results for question clarity, difficulty and life satisfaction.}
\label{tab:regressions} \\
  &\multicolumn{1}{c}{Questions clear}&\multicolumn{1}{c}{Questions easy}&\multicolumn{1}{c}{Life satisfaction}\\
\hline
\endfirsthead

\caption[]{Regression results for question clarity, difficulty and life satisfaction (cont.)} \\
  &\multicolumn{1}{c}{Questions clear}&\multicolumn{1}{c}{Questions easy}&\multicolumn{1}{c}{Life satisfaction}\\
\hline
\endhead

\multicolumn{4}{c}{\textit{Continued on next page}} \\
\endfoot

\hline
R2      &  0.09   &  0.06   &  0.22   \\
Observations  &  2068   &  2068   &  2068   \\
\endlastfoot

\textit{Gender:}    &                     &                     &                     \\
Male                &        0.00         &        0.00         &        0.00         \\
                    &         (.)         &         (.)         &         (.)         \\
Female              &        0.06         &       -0.04         &        0.09         \\
                    &     (0.069)         &     (0.063)         &     (0.095)         \\
\textit{Age group:} &                     &                     &                     \\
18-34               &        0.00         &        0.00         &        0.00         \\
                    &         (.)         &         (.)         &         (.)         \\
35-54               &       -0.09         &       -0.16         &       -0.33\sym{*}  \\
                    &     (0.092)         &     (0.087)         &     (0.131)         \\
55+                 &        0.13         &        0.10         &        0.16         \\
                    &     (0.130)         &     (0.117)         &     (0.176)         \\
\textit{Ethnicity:} &                     &                     &                     \\
White               &        0.00         &        0.00         &        0.00         \\
                    &         (.)         &         (.)         &         (.)         \\
Ethnic minority     &       -0.22\sym{*}  &       -0.08         &       -0.04         \\
                    &     (0.114)         &     (0.102)         &     (0.176)         \\
\textit{Marital status:}&                     &                     &                     \\
Married / Partnered &        0.00         &        0.00         &        0.00         \\
                    &         (.)         &         (.)         &         (.)         \\
Divorced / Separated&       -0.09         &       -0.04         &       -0.80\sym{***}\\
                    &     (0.099)         &     (0.092)         &     (0.139)         \\
Single              &       -0.01         &        0.05         &       -0.73\sym{***}\\
                    &     (0.154)         &     (0.149)         &     (0.210)         \\
Widowed             &       -0.07         &        0.26         &       -1.26\sym{***}\\
                    &     (0.196)         &     (0.170)         &     (0.298)         \\
Missing             &        0.14         &        0.34         &        0.13         \\
                    &     (0.200)         &     (0.179)         &     (0.293)         \\
\textit{Religion:}  &                     &                     &                     \\
Christian           &        0.00         &        0.00         &        0.00         \\
                    &         (.)         &         (.)         &         (.)         \\
Other faiths        &        0.20         &       -0.08         &        0.13         \\
                    &     (0.152)         &     (0.136)         &     (0.217)         \\
No faith            &        0.05         &        0.01         &       -0.07         \\
                    &     (0.079)         &     (0.073)         &     (0.109)         \\
Missing             &       -0.06         &       -0.09         &       -0.09         \\
                    &     (0.136)         &     (0.118)         &     (0.209)         \\
\textit{Sexuality:} &                     &                     &                     \\
Heterosexual / Missing&        0.00         &        0.00         &        0.00         \\
                    &         (.)         &         (.)         &         (.)         \\
Non-heterosexual    &        0.05         &        0.10         &       -0.09         \\
                    &     (0.111)         &     (0.099)         &     (0.148)         \\
\textit{Work status:}&                     &                     &                     \\
Working (FT or PT)  &        0.00         &        0.00         &        0.00         \\
                    &         (.)         &         (.)         &         (.)         \\
Unemployed / Non-working&       -0.04         &       -0.12         &       -0.43\sym{*}  \\
                    &     (0.146)         &     (0.138)         &     (0.218)         \\
Retired             &        0.05         &       -0.09         &        0.91\sym{***}\\
                    &     (0.150)         &     (0.131)         &     (0.207)         \\
Student / Other     &        0.07         &        0.06         &       -0.05         \\
                    &     (0.157)         &     (0.150)         &     (0.237)         \\
\textit{Household income:}&                     &                     &                     \\
Up to £19,999 per year&        0.00         &        0.00         &        0.00         \\
                    &         (.)         &         (.)         &         (.)         \\
£20,000 to £39,999 per year&        0.06         &       -0.07         &        0.04         \\
                    &     (0.117)         &     (0.110)         &     (0.170)         \\
£40,000 to £59,999 per year&        0.01         &       -0.03         &        0.15         \\
                    &     (0.132)         &     (0.120)         &     (0.192)         \\
£60,000 per year or more&        0.15         &        0.08         &        0.47\sym{*}  \\
                    &     (0.134)         &     (0.126)         &     (0.195)         \\
Missing             &       -0.15         &       -0.27\sym{*}  &        0.11         \\
                    &     (0.119)         &     (0.109)         &     (0.177)         \\
\textit{Sector:}    &                     &                     &                     \\
Private sector      &        0.00         &        0.00         &        0.00         \\
                    &         (.)         &         (.)         &         (.)         \\
Public sector       &       -0.04         &       -0.12         &        0.11         \\
                    &     (0.090)         &     (0.082)         &     (0.119)         \\
Third/voluntary sector&       -0.04         &       -0.18         &       -0.47\sym{*}  \\
                    &     (0.148)         &     (0.130)         &     (0.209)         \\
Missing             &        0.11         &        0.11         &       -0.35\sym{*}  \\
                    &     (0.123)         &     (0.112)         &     (0.172)         \\
\textit{Household size:}&                     &                     &                     \\
1                   &        0.00         &        0.00         &        0.00         \\
                    &         (.)         &         (.)         &         (.)         \\
2                   &       -0.16         &       -0.19         &       -0.17         \\
                    &     (0.110)         &     (0.100)         &     (0.156)         \\
3                   &       -0.22         &       -0.19         &       -0.12         \\
                    &     (0.124)         &     (0.115)         &     (0.172)         \\
4 or more           &       -0.01         &        0.04         &        0.00         \\
                    &     (0.126)         &     (0.117)         &     (0.176)         \\
Missing             &       -0.21         &       -0.22         &       -0.48         \\
                    &     (0.292)         &     (0.260)         &     (0.376)         \\
\textit{IMD decile:}&                     &                     &                     \\
Decile 1 to 3 (most deprived)&        0.00         &        0.00         &        0.00         \\
                    &         (.)         &         (.)         &         (.)         \\
Decile 4 to 7 (middle)&        0.11         &        0.03         &        0.08         \\
                    &     (0.084)         &     (0.077)         &     (0.120)         \\
Decile 8 to 10 (least deprived)&        0.10         &       -0.01         &        0.09         \\
                    &     (0.091)         &     (0.085)         &     (0.125)         \\
\textit{Education:} &                     &                     &                     \\
Low                 &        0.00         &        0.00         &        0.00         \\
                    &         (.)         &         (.)         &         (.)         \\
Medium              &        0.28\sym{**} &        0.13         &       -0.07         \\
                    &     (0.102)         &     (0.092)         &     (0.139)         \\
High                &        0.69\sym{***}&        0.42\sym{***}&       -0.25         \\
                    &     (0.107)         &     (0.098)         &     (0.145)         \\
\textit{Media source:}&                     &                     &                     \\
A printed copy of a newspaper&        0.00         &        0.00         &        0.00         \\
                    &         (.)         &         (.)         &         (.)         \\
A newspaper's website&        0.48\sym{*}  &        0.16         &       -0.27         \\
                    &     (0.215)         &     (0.193)         &     (0.311)         \\
A news website not associated with a newspaper&        0.45\sym{*}  &        0.20         &       -0.39         \\
                    &     (0.215)         &     (0.189)         &     (0.299)         \\
A news app on a mobile or tablet device&        0.43\sym{*}  &        0.21         &       -0.16         \\
                    &     (0.203)         &     (0.179)         &     (0.281)         \\
Social network websites&        0.27         &        0.17         &       -0.36         \\
                    &     (0.206)         &     (0.182)         &     (0.290)         \\
Television          &        0.12         &       -0.07         &       -0.12         \\
                    &     (0.198)         &     (0.174)         &     (0.276)         \\
Radio               &        0.10         &        0.02         &        0.03         \\
                    &     (0.220)         &     (0.196)         &     (0.309)         \\
Missing             &        0.07         &       -0.03         &       -0.63\sym{*}  \\
                    &     (0.221)         &     (0.195)         &     (0.322)         \\
\textit{Social grade:}&                     &                     &                     \\
AB                  &        0.00         &        0.00         &        0.00         \\
                    &         (.)         &         (.)         &         (.)         \\
C1                  &        0.02         &       -0.05         &       -0.29\sym{*}  \\
                    &     (0.091)         &     (0.085)         &     (0.124)         \\
C2                  &       -0.08         &        0.04         &       -0.33\sym{*}  \\
                    &     (0.108)         &     (0.099)         &     (0.147)         \\
DE                  &       -0.01         &       -0.01         &       -0.41\sym{**} \\
                    &     (0.106)         &     (0.099)         &     (0.157)         \\
\textit{Nation:}    &                     &                     &                     \\
England             &        0.00         &        0.00         &        0.00         \\
                    &         (.)         &         (.)         &         (.)         \\
Wales               &       -0.25         &       -0.24         &        0.00         \\
                    &     (0.145)         &     (0.127)         &     (0.219)         \\
Scotland            &        0.07         &        0.15         &        0.41\sym{**} \\
                    &     (0.126)         &     (0.119)         &     (0.149)         \\
Northern Ireland    &        0.09         &        0.04         &        0.05         \\
                    &     (0.152)         &     (0.143)         &     (0.239)         \\
\textit{Health limits activities:}&                     &                     &                     \\
Limited at all      &        0.00         &        0.00         &        0.00         \\
                    &         (.)         &         (.)         &         (.)         \\
Not limited         &        0.01         &        0.01         &        1.02\sym{***}\\
                    &     (0.080)         &     (0.074)         &     (0.115)         \\
Missing             &        0.16         &        0.05         &        0.19         \\
                    &     (0.265)         &     (0.230)         &     (0.394)         \\
\textit{Tenure:}    &                     &                     &                     \\
Own                 &        0.00         &        0.00         &        0.00         \\
                    &         (.)         &         (.)         &         (.)         \\
Rent                &       -0.02         &       -0.01         &       -0.22         \\
                    &     (0.089)         &     (0.080)         &     (0.125)         \\
Other / Neither     &       -0.00         &       -0.06         &       -0.29         \\
                    &     (0.133)         &     (0.123)         &     (0.188)         \\
Missing             &        0.76\sym{***}&        0.25         &       -0.09         \\
                    &     (0.188)         &     (0.190)         &     (0.269)         \\
\textit{Financial giving:}&                     &                     &                     \\
No                  &        0.00         &        0.00         &        0.00         \\
                    &         (.)         &         (.)         &         (.)         \\
Yes                 &        0.25\sym{***}&        0.12         &        0.14         \\
                    &     (0.073)         &     (0.067)         &     (0.098)         \\
Missing             &       -0.28         &       -0.06         &       -0.12         \\
                    &     (0.253)         &     (0.219)         &     (0.247)         \\
\textit{Volunteering:}&                     &                     &                     \\
No                  &        0.00         &        0.00         &        0.00         \\
                    &         (.)         &         (.)         &         (.)         \\
Yes                 &        0.01         &       -0.13         &        0.33\sym{*}  \\
                    &     (0.106)         &     (0.102)         &     (0.143)         \\
Constant            &        2.34\sym{***}&        2.97\sym{***}&        6.27\sym{***}\\
                    &     (0.290)         &     (0.265)         &     (0.418)         \\
\bottomrule
\end{longtable}
}

{\footnotesize \textit{Note:} Standard errors in parentheses, * p$<$0.05, ** p$<$0.01, *** p$<$0.001}

\newpage
\section{Mean estimates}
\label{app:mean}

This appendix reports mean (OLS) estimates corresponding to the median regression results in the main text. Three patterns are worth noting before turning to the tables. First, the OLS estimates are uniformly higher than the median estimates by roughly 0.10 to 0.25 across cells, reflecting the right-skewed distribution of individual inequality aversion. Second, the heterogeneity pattern is qualitatively similar, with only two exceptions. The age gradient that is sharp in the median (0.55, 0.46, 0.36 across age groups) is essentially flat under OLS (0.67, 0.61, 0.61). And volunteering becomes marginally significant under OLS ($p = 0.015$) but is not under the median ($p = 0.613$).

\subsection*{Overall estimate}

The OLS estimate of $\alpha$ is 0.62 ($\text{se} = 0.021$), compared with the median estimate of 0.48 ($\text{se} = 0.013$) reported in the main text. Both estimates exclude the design-fixed anchor step (2→3) and use the same $7 \times N$ paired observations and YouGov post-stratification weights.

\subsection*{Marginal weights}

\begin{table}[H]
\centering
\caption{Mean marginal value of someone going up from life satisfaction at the level shown to the next higher level of wellbeing (from 2 to 3 = 100)}
\label{tab:mean_weights}
\begin{tabular}{lccccccccc}
 LS step & 2$\rightarrow$3 & 3$\rightarrow$4 & 4$\rightarrow$5 & 5$\rightarrow$6 & 6$\rightarrow$7 & 7$\rightarrow$8 & 8$\rightarrow$9 & 9$\rightarrow$10 \\ \hline
Mean & 100 & 76 & 74 & 70 & 65 & 58 & 51 & 46 \\
\end{tabular}
\end{table}

\subsection*{Inequality aversion by question comprehension (OLS)}

\begin{table}[H]
\centering
\caption{Inequality aversion by question comprehension (OLS)}
\label{tab:mean_comprehension}
\begin{tabular}{lcclcc}
\multicolumn{3}{c}{Did the questions make sense to you?} & \multicolumn{3}{c}{How easy or difficult it was to answer?} \\
Response & $\alpha$ & se & Response & $\alpha$ & se \\
\midrule
Not at all & $-$0.01 & 0.053 & Very difficult & 0.03 & 0.062 \\
2 & 0.46 & 0.046 & 2 & 0.59 & 0.047 \\
3 & 0.60 & 0.040 & 3 & 0.73 & 0.035 \\
4 & 0.96 & 0.044 & 4 & 0.87 & 0.043 \\
Completely & 0.98 & 0.038 & Very easy & 0.74 & 0.045 \\
\midrule
\multicolumn{6}{l}{Joint test of equality (Clarity): \textit{p-value} $< 0.001$} \\
\multicolumn{6}{l}{Joint test of equality (Difficulty): \textit{p-value} $< 0.001$}
\end{tabular}
\par
\begingroup
\footnotesize
\textit{Notes:} $\alpha$ estimated by OLS from Equation \ref{eq:2}.
\endgroup
\end{table}

\subsection*{Inequality aversion by life satisfaction (OLS)}

\begin{table}[H]
\centering
\caption{Inequality aversion by individual's life satisfaction (OLS)}
\label{tab:mean_lsat}
\begin{tabular}{lcc}
Life Satisfaction & $\alpha$ & se \\
\midrule
$\leq 2$ & 0.67 & 0.095 \\
3 & 0.80 & 0.083 \\
4 & 0.71 & 0.075 \\
5 & 0.56 & 0.058 \\
6 & 0.68 & 0.045 \\
7 & 0.73 & 0.039 \\
8 & 0.51 & 0.043 \\
9 & 0.46 & 0.127 \\
10 & 0.34 & 0.075 \\
\midrule
\multicolumn{3}{l}{Joint test of equality: \textit{p-value} $< 0.001$}
\end{tabular}
\par
\begingroup
\footnotesize
\textit{Notes:} $\alpha$ estimated by OLS from Equation \ref{eq:2}. Life satisfaction levels 0, 1 and 2 are merged into a single bottom category due to small cell sizes, matching the main text.
\endgroup
\end{table}

\subsection*{Inequality aversion by charitable giving and health (OLS)}

\begin{table}[H]
\centering
\caption{Inequality aversion by charitable giving and health (OLS)}
\label{tab:mean_group2}
\begin{tabular}{lclclc}
\multicolumn{2}{c}{Financial giving} & \multicolumn{2}{c}{Volunteering} & \multicolumn{2}{c}{Health} \\
Response & $\alpha$ (se) & Response & $\alpha$ (se) & Response & $\alpha$ (se) \\
\midrule
No & 0.54 (0.027) & No & 0.61 (0.022) & Limited & 0.67 (0.037) \\
Yes & 0.72 (0.031) & Yes & 0.76 (0.061) & Not limited & 0.61 (0.025) \\
\midrule
\multicolumn{6}{l}{Joint test \textit{p-values}: Financial giving $< 0.001$, Volunteering = 0.015, Health = 0.234}
\end{tabular}
\par
\begingroup
\footnotesize
\textit{Notes:} $\alpha$ estimated by OLS from Equation \ref{eq:2}.
\endgroup
\end{table}

\subsection*{Inequality aversion by demographic characteristics (OLS)}

\begin{table}[H]
\centering
\caption{Inequality aversion by demographic characteristics (OLS)}
\label{tab:mean_group3}
\begin{tabular}{lccccc}
\multicolumn{2}{c}{Age group} & \multicolumn{2}{c}{Gender} & \multicolumn{2}{c}{Marital status} \\
\cmidrule(lr){1-2}\cmidrule(lr){3-4}\cmidrule(lr){5-6}
Response & $\alpha$ (se) & Response & $\alpha$ (se) & Response & $\alpha$ (se) \\
\midrule
18-34 & 0.67 (0.036) & Male & 0.69 (0.031) & Married / Partnered & 0.64 (0.027) \\
35-54 & 0.61 (0.034) & Female & 0.57 (0.027) & Divorced / Separated & 0.65 (0.038) \\
55+ & 0.61 (0.036) & & & Single & 0.69 (0.092) \\
 & & & & Widowed & 0.37 (0.095) \\
\midrule
\multicolumn{6}{l}{Joint test \textit{p-values}: Age group = 0.428, Gender = 0.004, Marital status = 0.034}
\end{tabular}
\par
\begingroup
\footnotesize
\textit{Notes:} $\alpha$ estimated by OLS from Equation \ref{eq:2}.
\endgroup
\end{table}



\subsection*{Inequality aversion by household income and country (OLS)}

\begin{table}[H]
\centering
\caption{Inequality aversion by household income and country (OLS)}
\label{tab:mean_income_country}
\begin{tabular}{lccc}
\multicolumn{2}{c}{Household income} & \multicolumn{2}{c}{Country} \\
\cmidrule(lr){1-2}\cmidrule(lr){3-4}
Response & $\alpha$ (se) & Response & $\alpha$ (se) \\
\midrule
Up to \pounds19,999 & 0.60 (0.050) & England & 0.63 (0.023) \\
\pounds20,000--39,999 & 0.58 (0.038) & Wales & 0.66 (0.093) \\
\pounds40,000--59,999 & 0.70 (0.044) & Scotland & 0.57 (0.055) \\
\pounds60,000+ & 0.74 (0.051) & Northern Ireland & 0.64 (0.086) \\
\midrule
\multicolumn{4}{l}{Joint test \textit{p-values}: Income = 0.038, Country = 0.775}
\end{tabular}
\par
\begingroup
\footnotesize
\textit{Notes:} $\alpha$ estimated by OLS from Equation \ref{eq:2}.
\endgroup
\end{table}

\end{document}